\documentclass[aps,twocolumn,pre,showpacs,preprintnumbers,amssymb]{revtex4}

\usepackage{graphicx}
\usepackage{dcolumn}
\usepackage{bm}

\begin{document}

\preprint{hyperinfl-own.tex}

\title{Finite-time singularity in the evolution of hyperinflation
episodes}

\author{Martin A. Szybisz}

\affiliation{Departamento de Econom\'{\i}a, Facultad
de Ciencias Econ\'omicas, Universidad de Buenos Aires,\\
Av. C\'ordoba 2122, RA--1120 Buenos Aires, Argentina}
\affiliation{Departamento de Filosof\'{\i}a del Derecho, Facultad
de Derecho, Universidad de Buenos Aires,\\
Av. Figeroa Alcorta 2263, RA--1425 Buenos Aires, Argentina}

\author{Leszek Szybisz}
 \altaffiliation{Corresponding author}
 \email{szybisz@tandar.cnea.gov.ar}
\affiliation{Laboratorio TANDAR, Departamento de F\'{\i}sica,
Comisi\'on Nacional de Energ\'{\i}a At\'omica,\\
Av. del Libertador 8250, RA--1429 Buenos Aires, Argentina}
\affiliation{Departamento de F\'{\i}sica, Facultad de
Ciencias Exactas y Naturales,\\
Universidad de Buenos Aires, Ciudad Universitaria,
RA--1428 Buenos Aires, Argentina}
\affiliation{Consejo Nacional de Investigaciones
Cient\'{\i}ficas y T\'ecnicas,\\
Av. Rivadavia 1917, RA--1033 Buenos Aires, Argentina}

\date{\today}

\begin{abstract}
A model proposed by Sornette, Takayasu, and Zhou for describing
hyperinflation regimes based on adaptive expectations expressed in
terms of a power law which leads to a finite-time singularity is
revisited. It is suggested to express the price index evolution
explicitly in terms of the parameters introduced along the theoretical
formulation avoiding any combination of them used in the original
work. This procedure allows to study unambiguously the uncertainties
of such parameters when an error is assigned to the measurement of the
price index. In this way, it is possible to determine an uncertainty
in the critical time at which the singularity occurs. For this
purpose, Monte Carlo simulation techniques are applied. The
hyperinflation episodes of Peru (1969-90) and Weimar Germany (1920-3)
are reexamined. The first analyses performed within this framework of
the very extreme hyper-inflations occurred in Greece (1941-4) and
Yugoslavia (1991-4) are reported. The study of the hyperinflation
spiral experienced just nowadays in Zimbabwe predicts a
singularity, i.e., a complete economic crash within two years.
\end{abstract}


\maketitle

\section{Introduction}
\label{sec:introduction}

Since about one decade there is significant interest in applications
of physical methods in social and economical sciences
\cite{stauffer99,stanley99,sornette03b}. For example, it has been
found that the logarithmic change of the market price in the case of a
hyperinflation episode shows some universal characteristics similar to
those observed in physical systems. In such a regime the price index
increases more rapidly than a simple exponential law \cite{mizuno02}.
Moreover, it has been shown that such a super-exponential law indeed
finishes with a finite-time singularity \cite{sornette03} like several
physical systems.

Let us recall that the rate of inflation $i(t)$ is defined as
\begin{equation}
i(t) = \frac{P(t)-P(t-\Delta t)}{P(t-\Delta t)}
= \frac{P(t)}{P(t-\Delta t)} - 1  \:, \label{infla}
\end{equation}
where $P(t)$ is the price at time $t$ and $\Delta t$ is the period of
the measurements. In economics the terminology ``hyperinflation''
is used in rather rough sense to specify very hight inflation that
is ``out of control'', a condition in which prices increase rapidly
as a currency loses its property as medium of exchange, store of
value, and unit of account. No precise definition of hyperinflation
is universally accepted. One simple definition requires a monthly
inflation rate of 20 or $30\%$ or more. In informal usage the term
is often applied to much lower rates. In 1956, Cagan published {\it
The Monetary Dynamics of Hyperinflation} \cite{cagan56}, generally
regarded as the first serious study of hyperinflation and its effects.
There it is defined that ``inflation rates per month exceeding
$50\%$'' determine a scenario of hyperinflation.

During periods of very hight inflation the frequent change of prices
destroys rapidly the real wages and the unknown future of the economic
structure diminishes the flow of inversions. Such situations are very
costly to society, the workers have to be paid more frequently (even
daily) and there are rushes to spend the currency before prices rise
further causing enormous ``{\it shoe-leather costs}'' (in economics it
means: resources wasted when inflation encourages people to do more
trips to banks and stores wearing out their shoes). These effects are
accompanied with a strong devaluation of the currency which causes a
decline in real public revenues increasing the fiscal deficit.
Hyperinflation reduces real value of taxes collected, which are often
set in nominal terms and by the time they are paid, real value has
fallen. This feature is known as the Olivera-Tanzi effect, after
Olivera \cite{olivera67} and Tanzi \cite{tanzi77} who were the first
to interpret it by means of standard analytical tools
\cite{canavese92}. The occurrence of the Olivera-Tanzi effect may
impulse a rapid expansion of nominal money and credit. If people
expect a money supply growth, then this would lead to expecting
higher inflation. The expectation of higher inflation raises inflation
rate even if money growth does not actually increase. Once people
start to expect an inflation regime, their expectations may lead to
strong positive feedbacks that make inflation run away. This scenario
produce an important crisis in the population. The real investment,
loans and development diminish, while the unemployment and political
unrest grow significantly. Moreover, some of these effects usually
continue after the hyperinflation has been stopped. Therefore, models
of hyperinflation are considered very useful by macro-economists
because detecting hyperinflation in an early stage might contribute
to avoid such a tragedy.

There are several remarkable historical examples of hyperinflation.
Some of them were originated after World Wars \cite{cagan56}. The
most famous and studied cases are those occurred in Germany and
Hungary in the early 1920's after World War I and in Hungary at the
end of World War II. However, there are other extreme examples, for
instance, that of Greece during the military occupation ($1941-4$)
\cite{cagan56,palairet00,lykogiannis02}. On the other hand, there
were other cases which are not related to wars like that in Latin
America in the 1980's \cite{imf,sargent06} and more recently that
occurred during the transition period from a Centrally Planned economy
to a more Free Market oriented economy in countries of East Europe and
in the successor states emerged after the dissolution of the Soviet
Union \cite{imf}.

As a matter of fact, the end of Central Planning in Europe - whether
it came as a result of slow decay or of rapid collapse - was
frequently accompanied by bursts of high inflation. An important
example of such a behavior is Yugoslavia, where two periods of very
hight inflation in short time were experienced. The first had a long
build-up during the 1980's and peaked in 1989 reaching high, but not
very extreme inflation only briefly. The second one developed just
at the end of Centrally Planed economy and enhanced by the Civil War
is the worst episode of hyperinflation in History. This very severe
hyperinflation occurred in the period $1991-4$ \cite{petrovic00,%
nielsen04}.

The inflation rates at the end of extreme hyper-inflations reached
very impressive amounts per month. For instance, the ratios of price
indexes for the most severe known incidents of inflation quoted by
Cagan \cite{cagan56} are: in Germany in 1923
$P$(November-13th)/$P$(October-16th)=10$^{2.556}$, i.e., inflation
rose about $3.6 \times 10^4$ percent per month; in Greece after
liberation from the occupation by German troops in 1944 the ratio $P$
(November-10th)/$P$(October-31th)=10$^{5.932}$ means $8.55 \times
10^7$ percent per ten days and in Hungary after the end of World War
II in 1946 $P$(end-July)/$P$(end-June)=10$^{14.6226}$ corresponds to
$4.2 \times 10^{16}$ percent per month.
%More recently, Yugoslavia suffered $5 \times 10^{15}$
%percent inflation per month (prices double every 16 hours) between
%1 October 1993 and 24 January 1994.

Despite all the unpleasant experiences with hight inflation,
sometimes, it is still considered to be the apple of paradise. Since
2000 Zimbabwe exhibits an increasing high inflation rate \cite{zrb},
which is already the highest in the world. Indeed, the spiral of
price growth developed in that country is nowadays considered an
important episode of hyperinflation.

Since a long time ago it is known that in the case of moderate
inflation the prices exhibit an exponential growth \cite{cagan56}. A
recent analysis of the hyperinflation in Germany, Hungary, Brazil,
Israel, Nicaragua, Peru, and Bolivia performed by Mizuno, Takayasu,
and Takayasu \cite{mizuno02} indicated that the price indexes or
currency exchange rates of these countries grew according to a
double-exponential function $e^{b_1e^{b_2t}}$ of time (with $b_1,
b_2>0$). It was shown that this super-exponential growth can be
obtained from a nonlinear positive feedback process in which the past
market price growth influences the people's expected future price,
which itself impacts the {\it a posteriori} realized market price.
This process is fundamentally based on the mechanism of ``adaptive
inflationary expectation'' and it is similar to the positive feedbacks
occurring during transmission of information due to imitative and herd
behaviors \cite{sornette03b,zimmermann00}.

The double-exponential model of Mizuno {\it et al.} \cite{mizuno02}
gives a useful mathematical description of hyperinflation, however,
it does not provide a rigorous determination of the end of the
hyperinflation regime. More recently, Sornette, Takayasu, and Zhou
\cite{sornette03} re-examined the theory developed in Ref.\
\cite{mizuno02} and showed that the double-exponential law is in fact
a discrete-time approximation of a general power law growth endowed
with a finite-time singularity at some critical time $t_c$.

Let us notice that singularities occur in different sorts of dynamical
systems and are spontaneously reached in finite time. Such a behavior
can be found in models of either physical or other kind of systems.
In the case of physics we can mention the Euler equations of inviscid
fluids \cite{bhattach95}, the surface curvature on the free surface
of a conducting fluid in an electric field \cite{zubarev98}, the
equations of General Relativity coupled to a mass field leading to
the formation of black holes \cite{choptuik99}, the vortex collapse
of systems of point vortexes \cite{leoncini00}, or the Euler's disks
as a rotating coin \cite{mocatt00}. On the other hand, this kind of
singularity is also present in models of micro-organisms aggregating
to form fruiting bodies \cite{rascle95} and in the dynamics of the
world population, and the economic and financial indexes
\cite{johansen00}.

The analysis of the finite-time singularity proposed in Ref.\
\cite{sornette03} allows to determine the theoretical end of the
hyperinflation regime. However, from the practical point of view it
is also important to estimate an uncertainty of $t_c$ in terms of
variations of $P(t)$. An analysis of this issue is just developed in
the present work.

In Sec.\ \ref{sec:theory} we revise the theoretical formulations
published in Refs.\ \cite{mizuno02,sornette03} introducing a careful
treatment of the initial time of the series of data $t_0$. The price
index evolution is expressed explicitly in terms of the free
parameters introduced along the theoretical formulation avoiding any
combination of them. In addition, we propose a procedure to determine
the uncertainties of the fitting parameters when an error is assigned
to the measurement of the price index. Monte Carlo simulation
techniques are used for the error analysis. In this way quotes for
$t_c$ are set. A study of the hyperinflation in Peru and Germany are
presented as testing cases. In Sec.\ \ref{sec:results} we report the
first studies of the very extreme hyper-inflations of Greece and
Yugoslavia performed within the framework outlined in the present
work. Furthermore, on the basis of an analysis of data of the current
trend of the hyperinflation in Zimbabwe, we predict an economic crash
in this country in about two years. Finally, the main conclusions are
summarized in Sec.\ \ref{sec:summary}.

\section{Theoretical background}
\label{sec:theory}

In the academic financial literature, the simplest and most robust
way to account for inflation is to take logarithm. Therefore, the
continuous rate of change in prices is usually defined as
\begin{equation}
C(t) = \frac{\partial \ln{P(t)}}{\partial t} \:. \label{c_rate0}
\end{equation}
Usually the derivative of Eq.\ (\ref{c_rate0}) is expressed in a
discrete way as
\begin{equation}
C(t) = \frac{\left[ \ln P(t+\Delta t)-\ln P(t) \right]}{\Delta t}
= \frac{1}{\Delta t}\,\ln \left[ \frac{P(t+\Delta t)}{P(t)} \right]
\:. \label{c_rate1}
\end{equation}
The growth rate of price over one period is defined as
\begin{equation}
r(t) \equiv C(t)\,\Delta t = \ln \left[ \frac{P(t+\Delta t)}{P(t)}
\right] = p(t+\Delta t) - p(t) \:. \label{rate1}
\end{equation}
Here, a notation widely utilized in the academic literature, $p(t) =
\ln P(t)$, is introduced. When $r(t)$ takes big values over a large
period of time a hyperinflation regime is to be reached.

In his pioneering work, Cagan \cite{cagan56} proposed a model of
inflation based on the mechanism of ``adaptive inflationary
expectation'' of positive feedback between realized growth of the
market price $P(t)$ and the growth of people's averaged expectation
price $P^*(t)$. These two prices are thought to evolve due to a
positive feedback mechanism: an upward change of market price $P(t)$
in a unit time $\Delta t$ induces a rise in the people's expectation
price $P^*(t)$, and such an anticipation pulls up the market price.

Cagan's assumption that the growth rate of $P^*(t)$ is proportional
to the past realized growth rate of the market price $P(t)$ is
expressed by the following equation
\begin{equation}
\frac{P(t+\Delta t)}{P(t)} = \frac{P^*(t)}{P(t)}
= \frac{P^*(t)}{P^*(t-\Delta t)} \:, \label{cag1}
\end{equation}
\begin{equation}
\frac{P^*(t+\Delta t)}{P^*(t)} = \frac{P(t)}{P(t-\Delta t)} \:.
\label{cag2}
\end{equation}
Now, one may introduce
\begin{equation}
r^*(t) \equiv  C^*(t)\,\Delta t = \ln \left[ \frac{P^*(t+\Delta t)}
{P^*(t)} \right] \:. \label{rate2}
\end{equation}
So, expressions (\ref{cag1}) and (\ref{cag2}) are equivalent to
\begin{equation}
r(t) = r^*(t-\Delta t) \:, \label{rate3}
\end{equation}
\begin{equation}
r^*(t) = r(t-\Delta t) \:, \label{rate4}
\end{equation}
whose solution is $r(t+\Delta t)=r(t-\Delta t)$ which indicates a
constant finite growth rate equal to its initial value $r(t)=r(t_0)
=r_0$. Since the market price is given by
\begin{equation}
P(t) = P(t_0)\,\exp{\left[\frac{1}{\Delta t}\int^t_{t_0} r(t') dt'
\right]} \;, \label{pt}
\end{equation}
Eqs.\ (\ref{rate3}) and (\ref{rate4}) lead to a steady state
exponential inflation
\begin{equation}
P(t) = P_0\,\exp{\left[\frac{r_0}{\Delta t}\,(t-t_0) \right]}
\;, \label{pt0}
\end{equation}
where $P(t_0)=P_0$. This form can be reduced to a linear form in $t$
\begin{equation}
\ln P(t) = \ln P_0 + C_0\,(t-t_0) \;,
\label{lpt0}
\end{equation}
where
\begin{equation}
C_0 = \frac{r_0}{\Delta t} \;, \label{C0}
\end{equation}
is the initial growth in prices.

\subsection{Double-exponential growth}
\label{sec:double}

Mizuno {\it et al.} \cite{mizuno02} have analyzed the
hyperinflation of Germany ($1920-3$), Hungary ($1945-6$), Brazil
($1969-94$), Israel ($1969-85$), Nicaragua ($1969-91$), Peru
($1969-90$) and Bolivia ($1969-85$), and showed that the price
indexes or currency exchange rates of these countries grew
super-exponentially according to a double-exponential function
$e^{b_1e^{b_2t}}$ of time (with $b_1, b_2>0$). These authors
generalized Eq.\ (\ref{rate2}) writing
\begin{equation}
\frac{P^*(t+\Delta t)}{P^*(t)} = \left(\frac{P(t)}{P(t-\Delta t)}
\right)^b = \left[1 + i(t)\right]^b\:, \label{cag21}
\end{equation}
which can be expressed as
\begin{equation}
r^*(t) = b\,r(t-\Delta t) \:. \label{rate41}
\end{equation}
Cagan's original model is recovered for the special case $b=1$. An
exponent $b$ larger than $1$ avoids systematic errors of other models
capturing the fact that the adjustment of the expected price $P^*(t)$
is weak for small changes of the realized market prices and becomes
very strong for large deviations.

The system of Eqs. (\ref{rate3}) and (\ref{rate41}) gives
\begin{equation}
r(t+\Delta t)=b\,r(t-\Delta t) \:. \label{miz0}
\end{equation}
In the continuous limit it becomes
\begin{equation}
\frac{dr}{dt} = b_2\,r(t) \;, \label{miz1}
\end{equation}
with $b_2=(b-1)/(2\,\Delta t) > 0$. The solution is
\begin{equation}
r(t) = r_0 e^{b_2\,(t-t_0)} \;. \label{drate}
\end{equation}
Upon introducing this result into Eq.\ (\ref{pt}) one gets the double
exponential form for the market price
\begin{equation}
P(t) = P_0\,\exp{\left\{\frac{C_0}{b_2}
\left[e^{b_2\,(t-t_0)}-1 \right]\right\}} \;. \label{d-exp}
\end{equation}
A straightforward calculation shows that in the limit $b_2 \to 0$ the
simple exponential of Eq. (\ref{pt0}) is recovered. For $b_2(t-t_0)
>> 1$ one gets the expression of Ref.\ \cite{mizuno02}
\begin{equation}
P(t) \propto e^{b_1\,e^{b_2\,(t-t_0)}} \:, \label{d-exp2}
\end{equation}
with $b_1=C_0/b_2$.

\subsection{Finite-time singularity}
\label{sec:finite}

A further generalization of the Cagan's model has been reported by
Sornette {\it et al.} \cite{sornette03}. These authors proposed a
different version of the nonlinear feedback process. They kept
expression (\ref{cag1}) or equivalently Eq.\ (\ref{rate3}) and
replaced Eq.\ (\ref{cag21}) or equivalently expression (\ref{rate41})
by
\begin{equation}
r^*(t) = r(t-\Delta t) + a\,[r(t-\Delta t)]^\gamma \:,~~~~~~with
~~\gamma > 1 \;. \label{rate42}
\end{equation}
Note that this formulation (\ref{rate42}) retrieves both previous
proposals: the Cagan's formulation (\ref{rate4}) is get for $a=0$
and the Mizuno {\it et al.}'s form (\ref{rate41}) is obtained for
$\gamma=1$.

The authors of Ref.\ \cite{sornette03} claim that their formulation
better captures the intrinsically nonlinear process of the formation
of expectations. Indeed, if $r(t-\Delta t)$ is small (explicitly, if
is is smaller $1/a^{1/\gamma}$), the second nonlinear term $a\,[r(t-
\Delta t)]^\gamma$ in the right-hand-side of (\ref{rate42}) is
negligible compared with the first Cagan's term $r(t-\Delta t)$ and
one recovers the exponentially growing inflation regime of normal
times. However, when the realized growth rate becomes significant,
people's expectations start to amplify these realized growth rates,
leading to a super-exponential growth. Geometrically, the difference
between the formulation of Eq.\ $(\ref{rate42})$ and that of Eq.\
$(\ref{rate41})$ consists in replacing a straight of slope $b$ larger
than $1$ by a upwards convex function with slope at the origin and
whose local slope increases monotonically with the argument.

This theory provides the first practical approach for predicting its
future path until its end. In practice, the end of an hyperinflation
regime is expected to occur somewhat earlier than at the asymptotic
critical time $t_c$, because governments and central banks are forced
to do something before the infinity is reached in finite time. Such
actions are the equivalent of finite-size and boundary condition
effects in physical systems undergoing similar finite-time
singularities. Hyperinflation regimes are of special interest as they
emphasize in an almost pure way the impact of collective behavior of
people interacting through their expectations.

\subsubsection{Determination of the critical time}
\label{sec:forms}

Putting Eq. (\ref{rate3}) together with expression (\ref{rate42})
leads to
\begin{equation}
r(t+\Delta t) = r(t-\Delta t) + a\,[r(t-\Delta t)]^\gamma \;.
\label{rate43}
\end{equation}
Taking the continuous limit, expression (\ref{rate43}) becomes
\begin{equation}
\frac{dr}{dt} = a_1\,[r(t)]^\gamma \;, \label{rate44}
\end{equation}
where $a_1$ is a positive coefficient with dimensions of the
inverse of time. In this case the growth rate accelerates with time
according to $[r(t)]^{\gamma-1}$. As a consequence of this power law
acceleration the solution of Eq.\ (\ref{rate44}) exhibits
singularities in finite-time \cite{bender78}
\begin{equation}
r(t) = r_0\,\left(\frac{t_c-t_0}{t_c-t}\right)^{1/(\gamma-1)} \;.
\label{rate45}
\end{equation}
The critical time is determined by the initial condition $r_0$, the
exponent $\gamma$, and the coefficient $a_1$
\begin{equation}
t_c = t_0 + \frac{1}{a_1\,(\gamma-1)\,[r_0]^{(\gamma-1)}}
\;. \label{c_time}
\end{equation}
We must notice that in Ref.\ \cite{sornette03} there are misprints:
i) the coefficient $a_1$ should be dropped from Eq.\ (14) and ii)
the expression for $t_c$ given just below that equation is incorrect.
See, for instance, Eq.\ (17) in Ref.\ \cite{ike02}.

A power law singularity is essentially indistinguishable from an
exponential of an exponential of time, except when the distance
$t_c-t$ from the finite time singularity becomes comparable with
the time step $\Delta t$. The main difference between the formulations
proposed by Mizuno {\it et al.} \cite{mizuno02} and Sornette {\it et
al.} \cite{sornette03} is that the latter one contains an information
on the end of the growth phase, embodied in the existence of the
critical $t_c$.

In this case, Eq.\ (\ref{pt}) leads to
\begin{eqnarray}
\ln \left[\frac{P(t)}{P_0}\right]&=&\frac{1}{\Delta t}\,\int^t_{t_0}
r(t') dt' \nonumber\\
&=& \frac{r_0}{\alpha}\,\left(\frac{t_c-t_0}{\Delta t}\right)\,
\left[\left(\frac{t_c-t}{t_c-t_0}\right)^{-\alpha} - 1\,\right]
\nonumber\\
&=& C_0\,\left(\frac{t_c-t_0}{\alpha}\right)\,
\left[\left(\frac{t_c-t}{t_c-t_0}\right)^{-\alpha} - 1 \,\right] \;,
\nonumber\\ \label{lpt}
\end{eqnarray}
with
\begin{equation}
\alpha = \frac{2-\gamma}{\gamma-1} \;. \label{alfa}
\end{equation}
For $t-t_0=\Delta t << t_c-t_0$ one obtains $\ln [P(t_0+\Delta t)/P_0]
=C_0\,\Delta t=r_0$ recovering the definition of Eq.\ (\ref{rate1}).

The time dependence of the market price $P(t)$ exhibits the two
different regimes depending on the sign of $\alpha$:

(i) For $1<\gamma<2$ one gets $\alpha > 0$ yielding a finite-time
singularity in the market price itself
\begin{eqnarray}
\ln \left[\frac{P(t)}{P_0}\right] = \frac{C_0\,(t_c-t_0)}{\alpha}\,
\left[\left(\frac{t_c-t_0}{t_c-t}\right)^{\alpha} - 1 \,\right]
\;. \label{price1}
\end{eqnarray}
In this regime the price exhibits a finite-time singularity at the
same critical value $t_c$ as the growth rate. Hence, this solution
corresponds to a genuine divergence of $\ln{P(t)}$. 

(ii) For $\gamma>2$ one gets $\alpha < 0$ yielding a finite-time
singularity in $r(t)$ but the market price evolve as
\begin{eqnarray}
\ln \left[\frac{P(t)}{P_0}\right] &=& \frac{C_0\,(t_c-t_0)}{\alpha}\,
\left[\left(\frac{t_c-t}{t_c-t_0}\right)^{-\alpha} - 1\,\right]
\nonumber\\
&=& \frac{C_0\,(t_c-t_0)}{\alpha'}\,\left[\,1 -
\left(\frac{t_c-t}{t_c-t_0}\right)^{\alpha'}\,\right]
\;, \label{price2}
\end{eqnarray}
with
\begin{equation}
\alpha' = -\alpha = \frac{\gamma-2}{\gamma-1} \;. \label{alfap}
\end{equation}
Here it holds $0<\alpha'<1$. As time approaches the critical value
$t_c$ the price converges to the value $C_0(t_c-t_0)/\alpha'$ leading
to equilibrium, this limit is reached more rapidly when $\alpha' \to
1$.

Let us now discuss the structure of Eq.\ (\ref{price1}) written in
the form
\begin{equation}
p(t) = p_0 + \frac{C_0\,(t_c-t_0)}{\alpha}\,
\left[\left(\frac{t_c-t_0}{t_c-t}\right)^{\alpha} - 1 \,\right]
\;. \label{price11}
\end{equation}
By fitting the measured price index to this expression one can
determine the critical time $t_c$ together with the parameters
$\alpha$ and $C_0$. However, since at $t=t_0$ the square bracket
vanishes, by using explicitly the assumed (measured) $P_0$, i.e.
setting $p_0=\ln P_0$, one would perform a fit with a fixed point at
the beginning of the hyperinflation episode which may bias the fitting
procedure for $t_c$. Therefore, it is convenient to consider $p_0$ as
an additional free parameter. On the oder hand, Eq.\ (\ref{price11})
leads to Eq.\ (15) of Ref.\ \cite{sornette03}
\begin{eqnarray}
p(t) &=& p_0 - \frac{C_0\,(t_c-t_0)}{\alpha} 
+ \frac{C_0\,(t_c-t_0)^{1+\alpha}}{\alpha\,(t_c-t)^\alpha}
\nonumber\\
&=& A + B(t_c-t)^{-\alpha} \;, \label{price12}
\end{eqnarray}
where
\begin{equation}
A = p_0 - \frac{C_0\,(t_c-t_0)}{\alpha} \;, \label{Asor}
\end{equation}
and
\begin{equation}
B = \frac{C_0\,(t_c-t_0)^{1+\alpha}}{\alpha} \;. \label{Bsor}
\end{equation}

Notice that in the formulation proposed in the present work, see Eq.\
(\ref{price11}), all the free parameters have their own physical
meaning: $p_0$ is the logarithm of $P_0$; $C_0$ is the initial growth
in price; $\alpha$ is fixed by the exponent $\gamma$ of the power law;
and $t_c$ is the end-point time of hyperinflation. While in the case
of Eq.\ (15) of Sornette {\it et al.} \cite{sornette03} the
coefficients $A$ and $B$ are combinations of that parameters. This
fact becomes important for an error analysis.

To compare episodes of hyperinflation it is useful to determine the
time interval, $\tau_2(t)$, needed for doubling the price index close
to the singularity. Starting from Eq.\ (\ref{rate45}) or Eq.\
(\ref{price1}) one can demonstrate that such a quantity is given by
\begin{equation}
\tau_2(t) = \frac{\ln 2}{C_0}\,
\left(\frac{t_c-t}{t_c-t_0}\right)^{1+\alpha}
= \frac{\ln 2}{\alpha\,B}\,(t_c-t)^{1+\alpha} \;. \label{tau2}
\end{equation}

\subsubsection{Estimation of uncertainties}
\label{sec:uncertainty}

Before analyzing hyperinflation episodes, we shall focus attention on
an important issue. Indeed, the determination of the fitting
parameters cannot be considered as unambiguous. Therefore, besides
giving the critical time $t_c$ and other parameters, it is also of
interest to provide an estimation of its uncertainties. Hence, in the
present work we analyzed the variation of the free parameters when
one takes into account an uncertainty in the measured rate of
inflation $i(t)$. The tabulated price index at a given time $t_n =
t_0+n\Delta t$ is evaluated according to a formula derived from Eq.\
(\ref{infla})
\begin{equation}
P(t=t_n) = \prod^n_{k=0} [1+i(t_k=k\Delta t)] \;, \label{Pmea}
\end{equation}
where $P_0=1+i(t_0)$ is usually fixed at unity, i.e., $i(t_0)$ is set 
zero. By assuming that each measured $i(t_k)$ has an uncertainty
$\Delta i(t_k)$ it is possible to assign an uncertainty to $P(t)$.
However, instead of evaluating $\Delta P(t)$, we preferred to estimate
directly the error of the fitting parameters.

For this purpose, we assumed that each value of the inflation rate
can be represented by a gaussian distribution with mean value
$\bar{\imath}_k=i(t_k)$ and a standard deviation $\sigma_k =
\Delta i(t_k)$. These distributions were sampled by using Monte Carlo
techniques in order to get for each $i(t_k)$ a random series of values
$i_j(t_k)$ with $j=1 \to m$. The size $m$ of these series was taken
large enough to satisfy to a good approximation
\begin{equation}
\bar{\imath}_k = \frac{1}{m} \sum^m_{j=1} i_j(t_k) = i(t_k)
\;, \label{x-mean}
\end{equation}
and
\begin{equation}
\bar{\sigma}_k = \sqrt{\frac{1}{m} \sum^m_{j=1}
[i_j(t_k)-\bar{x}_k]^2} = \Delta i(t_k) \;. \label{s-var}
\end{equation}
In this way $m$ generations of inflation rates were built, each one
is labeled by $j$ and composed of the obtained values $i_j(t_k)$ for
all $k$. Then each generation $j$ was fitted to Eq.\ (\ref{price11})
providing sets of fitting parameters $t_c(j)$, $\alpha(j)$, $C_0(j)$,
and $p_0(j)$. In order to speedup the procedure programs like those
included in Ref.\ \cite{bevington} may be used. Finally, for each
parameter the average value and variance were evaluated by using
expressions like that of Eqs.\ (\ref{x-mean}) and (\ref{s-var}). These
averages $\bar{t}_c$, $\bar\alpha$, $\bar{C}_0$, and $\bar{p}_0$ were
compared with the values $t_c$, $\alpha$, $C_0$, and $p_0$ yielded by
a direct fit of the tabulated $P(t)$, and the corresponding
differences were evaluated. When all the ratios of these differences
over the corresponding standard deviations were smaller that $0.1$
the obtained results have been accepted.

To perform this kind of error analysis in the case of a fit to Eq.\
(15) of Sornette {\it et al.} \cite{sornette03} would not be
appropriate because both coefficients $A$ and $B$ depend explicitly
on the remaining parameters $t_c$ and $\alpha$ and therefore their
errors would be strongly correlated.

\begin{table*}
\caption{\label{tab:table1} Analyzed hyper-inflations, the critical
time $t_c$ and remaining fitting parameters with the estimated
uncertainties.}
\begin{ruledtabular}
\begin{tabular}{lllllcccr}
Country & Currency & Period & \multicolumn{4}{c}{Parameters}
& $\chi$ & Ref. \\
&&& $t_c$ & $\alpha$ & $C_0$ & $p_0$ && \\
\hline
Peru\footnote{Inflation data are taken from a Table published by the
International Monetary Fund (IMF) \cite{imf}.} 
& Inti & 1969-1990 & $1991.29\pm0.37$ & $0.29\pm0.13$ & $0.18\pm0.02$
& $-0.38\pm0.07$ & 0.322 & PW \\
&&& 1991.29 & 0.3 &&& 0.291 & \cite{sornette03} \\
Zimbabwe\footnote{Inflation data are taken from Ref.\
\cite{zrb}.}
& ZW-Dollar & 1980-2007 & $2009.50\pm0.76$ & $0.79\pm0.21$ &
$0.08\pm0.01$ & $0.10\pm0.06$ & 0.234 & PW \\
Germany\footnote{Inflation data are taken from Ref.\ \cite{cagan56}.}
& Mark & 1920:01-1921:05 &&& $-0.008\pm0.004$ & $0.64\pm0.11$ & 0.076
& PW \\
&& 1921:05-1923:11 & 1924:01:05$\pm$11 & $0.56\pm0.12$
& $0.103\pm0.012$ & $0.57\pm0.09$ & 0.580 & PW \\
&& 1920:01-1923:11 & 1923:12:18 & 0.6 &&& 0.490
& \cite{sornette03} \\
Greece$^b$ & Drachma & 1941:04-1942:10 &&& $0.263\pm0.015$
& $0.14\pm0.09$ & 0.124 & PW \\
&& 1943:02-1944:10 & 1944:12:02$\pm$13 & $0.17\pm0.14$ 
& $0.210\pm0.022$ & $3.91\pm0.09$ &  0.230 & PW \\
Yugoslavia\footnote{Inflation data are taken from Ref.\
\cite{petrovic00}.}
& Dinar & 1990:12-1994:01 & 1994:03:10$\pm$4 & $0.53\pm0.05$ &
$0.335\pm0.018$ & $-1.52\pm0.14$ & 0.930 & PW \\
\end{tabular}
\end{ruledtabular}
\end{table*}

\begin{table}
\caption{\label{tab:table2} Parameters $A$ and $B$ and exponent
$\gamma$ of the power law given by Eq.\ (\ref{rate44}) together with
its uncertainty.}
\begin{ruledtabular}
\begin{tabular}{llrrcr}
Country & \multicolumn{4}{c}{Parameters}
& Ref. \\
& $\Delta t$ & $A$ & $B$ & $\gamma$ \\
\hline
Peru & year & -14.16 & 34. & $1.78\pm0.08$ & PW \\
     &      & -14.17 & 34. & 1.8 & \cite{sornette03} \\
     
Zimbabwe   & year  &&& $1.56\pm0.07$ & PW \\
Germany    & month &  -5.22 &  274.\footnote{For the evaluation of
this quantity time is taken in days.} & $1.64\pm0.05$ & PW \\
           &       &  -5.09 &  272.$^a$ & 1.6 & \cite{sornette03} \\
Hungary    &       &  -1.02 & 2370.$^a$ & 1.5 & \cite{sornette03} \\
Greece     & month & -21.62 &   78.$^a$ & $1.85\pm0.10$ & PW \\
Yugoslavia & month & -25.69 & 1030.$^a$ & $1.65\pm0.02$ & PW \\
\end{tabular}
\end{ruledtabular}
\end{table}

%\begin{figure*}
\begin{figure}
\centering\includegraphics[width=8cm, height=7cm]{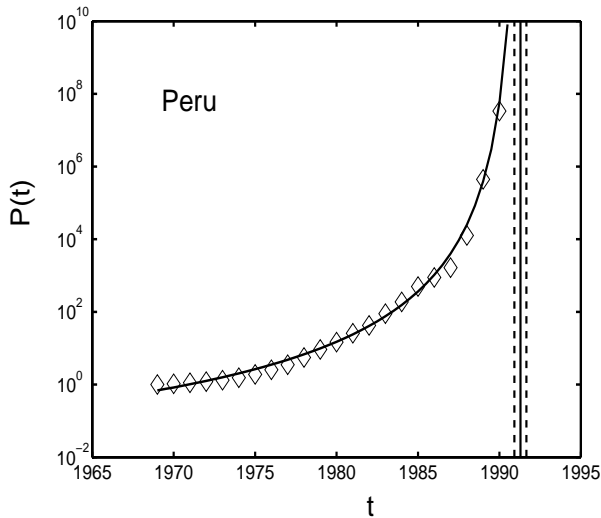}
\caption{\label{fig:Peru} Semi-logarithmic plot of the yearly
price index of Peru from 1969 to 1990 marked with open diamonds
[$P(t_0=1969)=1$] and its fit to Eq.\ (\ref{price11}). The vertical
solid line indicates the predicted critical time $t_c$ and vertical
dashed lines are error quotes for it, see text for explanations.}
\end{figure}
%\end{figure*}

\section{Analysis and numerical results}
\label{sec:results}

In a first step, before beginning the analysis of the episodes
announced in Sec.\ \ref{sec:introduction}, we shall describe the
application of our procedure to cases already treated by Sornette
{\it et al.} \cite{sornette03}. In particular, we shall report the
studies of the hyperinflation cycles of Peru and Weimar Germany. The
first case is a process developed over two decades, while the latter
one was build-up over a couple of years only. In this way the selected
checking examples cover the evolution characterizing the cases to be
analyzed for the first time within the framework outlined above.

Next, we shall analyze the very extreme cases of Greece and
Yugoslavia. Furthermore, we shall deal with the hyperinflation
exhibited nowadays by the economic system in Zimbabwe. This case is
very encouraging because the spiral of increasing prices is not
finished yet. Hence, an {\it a priori} prediction for the critical
time can be made.

At the end of the section we shall compare the most severe
hyper-inflations.

\subsection{Checking cases}
\label{sec:checking}

Figure \ref{fig:Peru} shows the price index of Peru during the period
1969-90. The parameters obtained from a mean-square fit of these data
to Eq.\ (\ref{price11}) together with the root-mean-square residue of
the fit, $\chi$, are quoted in Table \ref{tab:table1}. In order to
facilitate a quantitative comparison with the analysis of Sornette
{\it et al.} \cite{sornette03} their values of $t_c$ and $\alpha$ are
included in Table \ref{tab:table1}, while that of $A$ and $B$ are
quoted in Table \ref{tab:table2} together with our evaluations by
means of Eqs.\ (\ref{Asor}) and (\ref{Bsor}). A glance at these tables
indicates a perfect agreement between both fits. The quality of the
fit can be observed in Fig. \ref{fig:Peru}, this figure is to be
compared with Fig.\ 2 of Ref.\ \cite{sornette03}.
%Let us notice that a similar agreement was also obtained in the case
%of the hyperinflation episode of Bolivia (1969-85).

\begin{figure}
\centering\includegraphics[width=8cm, height=7cm]{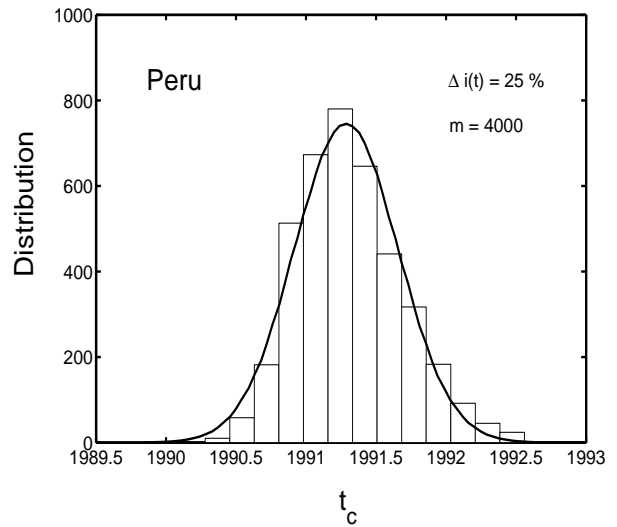}
\caption{\label{fig:distr} Distribution of the values of $t_c$
determined from solutions of Eq.\ (\ref{price11}) for the set of price
index obtained when the error assigned to measured inflation rate
$i(t)$ is $25\%$. The solid curve is the gaussian distribution.}
\end{figure}

\begin{figure}
\centering\includegraphics[width=8cm, height=7cm]{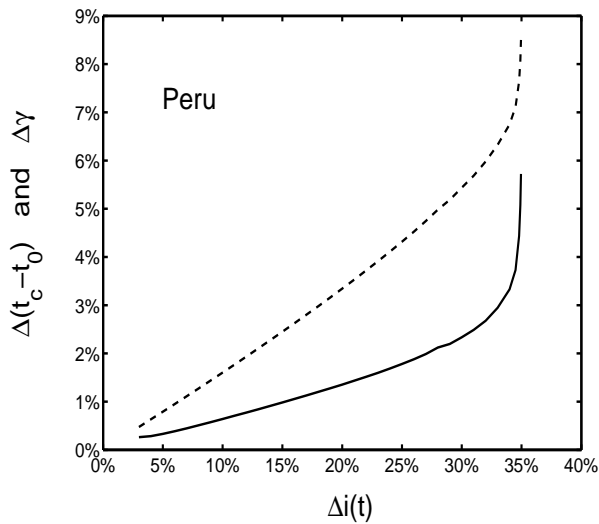}
\caption{\label{fig:error} Standard deviation (s.d.) for free
parameters obtained from solutions of Eq.\ (\ref{price11}) as a
function of the error assigned to measured inflation rate $i(t)$
expressed in percent. The solid and dashed curves are the s.d. for
$t_c-t_0$ and $\gamma$, respectively, also expressed in percent.}
\end{figure}

For the analysis of the uncertainties we assumed that all $i(t)$ have
the same relative error $\Delta i(t) [\%]$. The relative errors of
the parameters were determined as a function of $\Delta i(t)$
according the procedure outlined in Sec.\ \ref{sec:uncertainty}. The
requirements of Eqs.\ (\ref{x-mean}) and (\ref{s-var}) are reached
satisfactorily well for $m > 2000$. The distribution of results for
$t_c$ obtained by solving Eq.\ (\ref{price11}) for each generation of
$P(t)$ built up from Monte Carlo samplings at $\Delta i(t)=25\%$ is
displayed in Fig.\ \ref{fig:distr} together with the corresponding
gaussian distribution. This comparison shows a fair agreement. For
bigger $\Delta i(t)$ the distribution becomes asymmetric exhibiting a
repulsion towards larger $t_c$ and the standard deviation increases
dramatically. This effect is shown in Fig.\ \ref{fig:error}, where the
results for ``the most important parameters'', i.e., the critical
time $t_c-t_0$ and exponent $\gamma$ of the power law for the growth
rate $r(t)$
\begin{equation}
\gamma = \frac{2+\alpha}{1+\alpha} \label{gama}
\end{equation}
obtained for $3\% \lesssim \Delta i(t) \lesssim 35\%$ and $m =4000$
are plotted. These data indicate that the uncertainties of the
parameters are to a good approximation linear functions up to $\Delta
i(t) \approx 30\%$, then increase dramatically. At $\Delta i(t)
\approx 35\%$ a distribution equivalent to that displayed in Fig.\
\ref{fig:distr} differs from a gaussian. In this work we adopted
$25\%$ as a reasonable relative error of the measured $i(t)$. The
uncertainties of the fitting parameters quoted in Table
\ref{tab:table1} correspond to that quote. The uncertainty of $t_c$
is also displayed in Fig.\ \ref{fig:Peru}.

\begin{figure}
\centering\includegraphics[width=8cm, height=7cm]{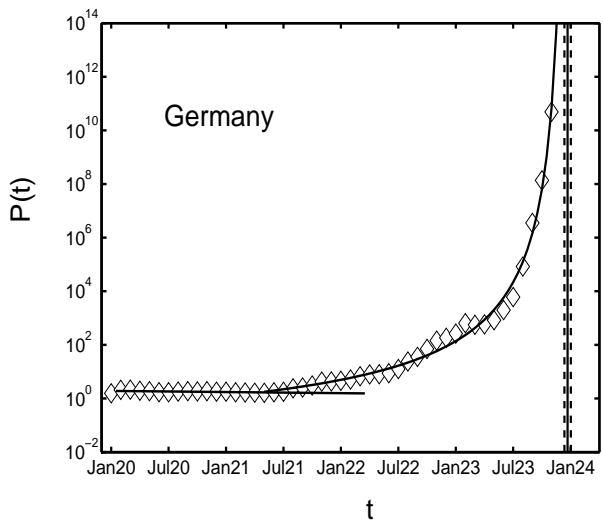}
\caption{\label{fig:Germany} Semi-logarithmic plot of the monthly
price index of Germany from 1920:01 to 1923:11 marked with open
diamonds. The straight line is a fit to Eq.\ (\ref{lpt1}) for the
period 1920:01 to 1921:05, while the solid curve is the fit to Eq.\
(\ref{price11}) for the period 1921:05 to 1923:11. The vertical solid
line indicates the predicted critical time $t_c$ and vertical dashed
lines are its error quotes, see text for explanations.}
\end{figure}

Sornette {\it et al.} \cite{sornette03} have analyzed the evolution of
the exchange rate between the German Mark and the US dollar during the
period 1920:01-1923:11 (from now on when dealing with monthly data the
notation Year:Month:Day will be used). The fit of all that data to
Eq.\ (\ref{price12}) yielded the values quoted in the present Tables
\ref{tab:table1} and \ref{tab:table2}. In order to compare these
results with that provided by an alternative information, we analyzed
the evolution of the price index taking the data from Table B3 of
Ref.\ \cite{cagan56}. Figure \ref{fig:Germany} shows the price index
of Germany during the period 1920:01-1923:11, the open diamonds are
values normalized to $P(t_0=1919:12:15)=1$. A comparison of this
figure with Fig. 4 of Ref.\ \cite{sornette03} indicates that the
cumulated price index over the considered period is similar to the
total variation of the exchange rate. It is important to point out
that data of the exchange rate were taken at the beginning of each
month, while the values of the price index correspond to the middle
of the month.

A careful analysis of price index indicates that during the period
1920:01 to 1921:05 the cumulated inflation is approximately zero.
This regime can be well described by the original theory of Cagan
condensed in Eq.\ (\ref{lpt0}), which can be cast into the form
\begin{equation}
p(t) = p_0 + C_0\,(t-t_0) \;. \label{lpt1}
\end{equation}
A fit to this equation yielded the values of $p_0$ and $C_0$ given
in Table \ref{tab:table1}, where the uncertainties corresponding to an
error of $25\%$ in the measured $i(t)$ are also quoted. The quality
of the fit may be observed in Fig.\ \ref{fig:Germany}. The slightly
negative slope $C_0$ is due to the fact that in this period several
months exhibit deflation. Consequently, in order to study the
hyperinflation episode we fitted to Eq.\ (\ref{price11}) data of the
period 1921:05 to 1923:11 only. This procedure yielded the values of
the free parameters and the $\chi$ listed in Table \ref{tab:table1}.
The quality of the fit may be observed in Fig.\ \ref{fig:Germany} and
it is similar to that obtained in Ref.\ \cite{sornette03} as indicated
by the values of $\chi$. The evaluated results of $A$ and $B$ are
included in Table \ref{tab:table2} (for the evaluation of these
quantities the time is taken in days). The agreement between the
values for $\alpha$, $A$, and $B$ obtained in the present work and
that of Ref.\ \cite{sornette03} is quite good. In the case of the
critical time there is a delay of about 15 days mainly caused by the
shift of the date attributed to measurements.

An error analysis similar to that described in the case of Peru
indicates that the uncertainties of the fitting parameters are linear
functions beyond $\Delta i(t)=35\%$. This is due to the fact that the
price index at the end of the period of measurement rose a bigger
value in the case of Germany ($P(t_{max}) \approx 5 \times 10^{10}$)
than of Peru ($P(t_{max}) \approx 3 \times 10^{7}$) and, therefore,
the finite-time singularity is better defined in the former case. The
photos of banknotes of hyperinflation episodes may be found in the
paper money gallery at the web site \cite{chao}.

For the hyperinflation regimes analyzed in the remaining part of this
section the error of the free parameters are determined assuming
$\Delta i(t)=25\%$. This uncertainty is large enough to provide a
``reasonable'' error quotes.

\subsection{The Greek catastrophic episode}
\label{sec:Greece}

Let us begin the description of the Greek hyperinflation by citing a
fragment of the talk addressed by Nicholas C. Garganas, Governor of
the Bank of Greece \cite{Garganas}. He said:
``In April 1941, the Axis Powers occupied Greece. For several years,
London became the seat of both the exiled Greek government and the
Bank of Greece, with the Bank's gold secretly transferred to South
Africa. Within occupied Greece , the economic situation became
increasingly grim and hundreds of thousands of Greeks died of hunger.
The Axis powers forced the country to pay not only for the upkeep
of the occupying troops, but also for their military operations in
Southeastern Europe. The puppet regime established by the occupiers
forced the Bank of Greece to resort to the printing press. As a
result, the country was beset with hyperinflation; between April 1941
and October 1944, the cost of living rose $2.3 \times 10^9$ times. In
these difficult circumstances, the country's economic system
collapsed. To give another example of the magnitude of inflation
during the occupation, let me mention that in November 1944,
immediately after liberation, a so-called ``new'' drachma was
introduced; it was set equal to $50 \times 10^9$ ``old'' drachmas!'' 

\begin{figure}
\centering\includegraphics[width=8cm, height=7cm]{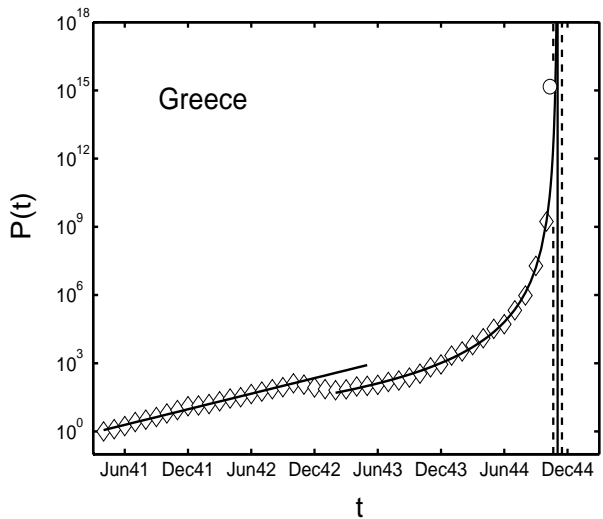}
\caption{\label{fig:Greece} Semi-logarithmic plot of the monthly
price index of Greece from 1941:04 to 1944:10 marked with open
diamonds. The open circle is the value of 1944:11. The straight line
is a fit to Eq.\ (\ref{lpt1}) for the period 1941:04 to 1942:10,
while the solid curve is a fit to Eq.\ (\ref{price11}) for the
period 1943:02 to 1944:10. The vertical solid line is the predicted
critical time $t_c$, while the vertical dashed lines indicate its
error bars.}
\end{figure}

Figure \ref{fig:Greece} shows in a semi-logarithmic plot the price
index of Greece taken from Table B6 of Ref.\ \cite{cagan56}. The open
diamonds are monthly data taken at the end of each month. The open
circle is the value for 1944:11:10, i.e., it corresponds to the first
10 days of November. These data are normalized to $P(t_0=1941:04:30)
= 1$. In this drawing one can observe the value $2.3 \times 10^9$
mentioned in the previous paragraph. A simple inspection indicates
two well differentiated regimes of inflation. This behavior can be
understood in terms of different phases of the foreign occupation.

At the beginning conquered Greece was divided into three zones of
control by the occupying powers, Germany, Italy and Bulgaria.
The Germans limited themselves during the first period of the
occupation to the strategically important areas as Athens, Central
Macedonia, Western Crete, and the islands of the Northern Aegean, and
their forces were limited. Bulgaria annexed Thrace and Eastern
Macedonia, while Italy occupied the greater part of the country.
Between the occupation zones no movements of goods and people was
allowed. The naval blockade coupled with transfers of agricultural
produce to Germany led to the gradual but firm establishment of a
black market.

Over the period 1941:04 to 1942:10, a dynamics of high inflation can
be observed. The cumulated inflation has been about $130\%$, hence,
the level of a hyperinflation was still not reached. This regime can
be well described by the original theory of Cagan given by Eq.\
(\ref{lpt1}). A fit of the price index to that expression yielded
the values of the parameters $p_0$ and $C_0$ quoted in Table
\ref{tab:table1}. The quality of the fit is quite good.

Until the summer of 1942 the resistance movement was in its infancy,
however, at the end of that year became strong. The spectacular
destruction of the Gorgopotamos bridge by a force of Greek guerrillas
and British saboteurs on 25 November caused a reaction of the Italian
authorities, in spite of it, the guerrillas were largely successful
in this region, creating ``liberated'' areas in the mountainous
interior including some towns.

This initial success of the guerrilla diminished the tension in the 
population causing a period of small deflation over four months. This
fact can be clearly seen in Fig.\ \ref{fig:Greece}.

However, the pressure of foreign troops increased and in 1943 German
elite troops were brought into the whole Greece. A heavy resistance
led to German contra-attacks and reprisals. In September 1943 the
Italians surrendered following the Allied invasion of Italy.
Throughout late 1943 and the first half of 1944, the Germans, in
cooperation with the Bulgarians and aided by Greek collaborators
launched clearing operations against the Greek resistance.

During this period, the German forced the Greek treasury to pay huge
amounts of ``occupation expenses''. Since the government of Greece
could not meet such an obligation from fiscal taxation (Olivera-Tanzi
effect) new money was printed (seigniorage). The attempt to control
prices failed and the fall off of production in devastated Greece's
economy led to the collapse of the normal markets and to an increase
of the black market.

Due to the general scenario of the war, the Germans were forced to
evacuate mainland Greece in October 1944. Their withdrew was finished
on November 2 and the exiled government returned to Athens. However,
already with the prospect of the liberation of Greece two resistance
groups (left and right orientated organizations) began to fight for
power. These tensions led almost immediately to a disastrous civil
war.

The very difficult situation of the later years of the occupation
caused the catastrophic hyperinflation \cite{cagan56,freris86}
displayed in Fig.\ \ref{fig:Greece}. The inflation reached a
peak in November 1944 after liberation. As mentioned in Sec.\
\ref{sec:introduction}, in the first ten days of November the
inflation rose the incredible value $8.55 \times 10^7\,\%$ (see Cagan
\cite{cagan56}). The monthly data from 1942:02 to 1944:10 were fitted
to Eq.\ (\ref{price11}). The results of the free parameters are listed
in Table \ref{tab:table1}, the quoted uncertainties correspond to an
error of $25\%$ in the measurements of inflation rates. The
uncertainty in $t_c$ is also displayed in Fig.\ \ref{fig:Greece}.

By looking at Fig.\ \ref{fig:Greece} one realizes that the solid
curve calculated with Eq.\ (\ref{price11}) reproduces very well the
measured data. Moreover, according to the solid curve the measured
value on November 10 would be reached on November 25. This anticipated
``explosion'' of the market price has been caused by the interplay of
different strongly increasing ``variables'', such as lack of goods and
political uncertainty mentioned above. The obtained critical time
predicts the definitive crash would be on 1944:12:02. The uncertainty
estimated by assuming that the inflation is measured with an error of
$25\%$ amounts about 13 days.

The stratification of wealth caused by hyperinflation and black
markets during the occupation seriously hindered postwar economic
development. The Greek government undertook several stabilization
efforts spread over a couple of years before price level stability
was achieved. These facts are described in the books written by
Palairet \cite{palairet00} and Lykogiannis \cite{lykogiannis02}. The
efforts to confront the hyperinflation consisted of a currency
conversion (convertibility of the new drachma into British Military
Authority Pounds), the creation of an independent supra-central bank
limiting the government's overdraft at the Bank of Greece, and a few
fiscal reforms to increase taxes or reduce expenditures
\cite{makinen84}.

\subsection{Yugoslavia - The worst episode in History}
\label{sec:Yugoslavia}

The residual Yugoslavia has experienced the highest recorded
hyperinflation in History. This episode occurred during a period of
two difficult circumstances like the transformation from a Centrally
Planned to a rather Free Market economy and the disastrous civil war
1991-4.

Let us now present a summary of the main events of the Yugoslav civil
war. A last effort to avoid Yugoslavia's disintegration was made 3
June 1991 through a joint proposal by Macedonia and Bosnia and
Herzegovina, offering to form a ``community of Yugoslav Republics''
with a centrally administered common market, foreign policy, and
national defense. However, Serbia opposed the proposal and the
former federal republic of Yugoslavia began the process of
dissolution. On 25 of June 1991 Slovenia and Croatia both declared
their independence from Yugoslavia. The national army of Yugoslavia,
then made up of Serbs controlled by the government of Belgrade,
stormed into Slovenia but failed to subdue the separatists there and
withdrew after only ten days of fighting losing interest in a
country with almost no Serbs. Instead, the attention was turned to
Croatia, a Catholic country where Orthodox Serbs made up 12 percent
of the population. The civil war started. In order to ``protect'' the
Serbian minority the central forces aided by Serbian guerrillas
invaded Croatia in July 1991. By the end of 1991, a U.S.-sponsored
cease-fire agreement was brokered between the Serbs and Croats
fighting in Croatia. Macedonia opted for independence on 20 November
1991. On 29 February 1992, the multi-ethnic republic of Bosnia and
Herzegovina passed a referendum for independence, but not all Bosnian
Serbs agreed. Under the guise of protecting the Serb minority in
Bosnia, the government of Belgrade channeled arms and military support
to them.

So, while the secession of Slovenia and Macedonia came relatively
peacefully, in the case of Croatia and Bosnia there were devastating
wars. In April 1992, the U.S. and European Community chose to
recognize the independence of Bosnia, a mostly Muslim country where
the Serb minority made up 32 percent of the population. On April 27,
Serbia joined the republic of Montenegro in a smaller New Yugoslav
Republic and responded to Bosnia's declaration of independence by
attacking their capital Sarajevo. Foreign governments responded with
sanctions, an embargo was introduced by the United Nations embargo on
30 May 1992.

\begin{figure}
\centering\includegraphics[width=8cm, height=7cm]{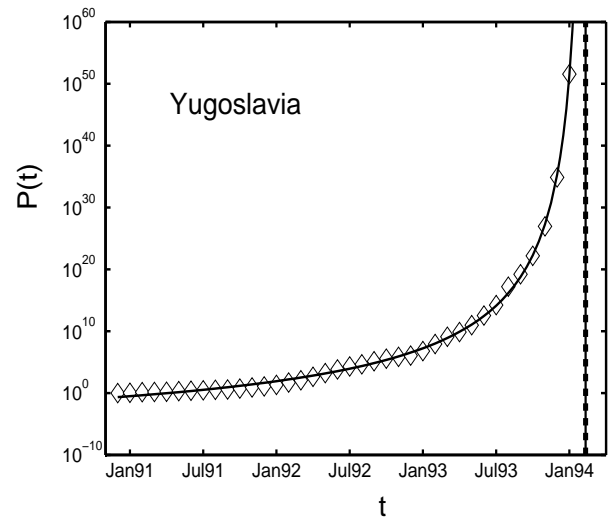}
\caption{\label{fig:Yugoslavia} Semi-logarithmic plot of the monthly
price index of Yugoslavia from 1990:12 to 1994:01 marked with open
diamonds. Data are normalized to $P(t_0$=1990:12)=1. The
solid curve is the fit to Eq. (\ref{price11}). The vertical lines
indicate predicted critical time $t_c$ and its error bars.}
\end{figure}

The disintegration of the former Yugoslavia led to decreased output
and fiscal revenue, while transfers to the Serbian population in
Croatia and Bosnia-Herzegovina as well as military expenditure added
to the fiscal problems. Simultaneously, the economy was changing
from Central Planning to Free Market. In order to finance the
increasing deficit, caused among others by the Olivera-Tanzi effect,
the government printed money. High inflation started to build up in
1991. The sources for the data of the complete cycle are documented
in Petrovi\'c and Mladenovi\'c \cite{petrovic00}. Figure
\ref{fig:Yugoslavia} shows the time series of monthly price index in
the period 1990:12 to 1994:01. The monthly inflation had already
risen the category of hyperinflation at the beginning of 1992 and
accelerated further despite the price freeze attempted in the end
of 1993:08. The overall impact of hyperinflation on the price index
reached about $10^{50}$ as it is also shown in Fig.\ 1 of
\cite{nielsen04}.

As a consequence of the hyperinflation Yugoslavia went through
several currency reforms simply removing zeros from the paper money.
In spite of these reforms, along this period the highest denomination
reached very large values. The largest nominal value ever officially
printed in Yugoslavia, a banknote of 500,000,000,000 (500 billion)
dinars was released at the end of 1993 (see photo in Ref.\
\cite{chao}). The overall impact of hyperinflation on currency at the
final reform for stabilization performed at the end of 1994:01 was:
1 Novi Dinar $= 1.2 \times 10^{27}$ pre 1990 Dinar, equivalent to the
cumulated price index until 1993:11 (see Fig.\ \ref{fig:Yugoslavia}).
%1,200,000,000,000,000,000,000,000,000 pre 1990 Dinara.

Many Yugoslavian businesses refused to take the Yugoslavian currency
at all and the German Deutsche Mark effectively became the currency
of Yugoslavia. But government organizations, government employees
and pensioners still got paid in Yugoslavian dinars so there was
still an active exchange in dinars. However, farmers selling in the
free markets refused to sell food for Yugoslavian dinars. So, many
monetary transactions were actually taking place in German Marks
rather than local currency. Therefore, the hyperinflation can be also
measured in terms of the black market exchange rate for German Marks
and Yugoslavian Dinars. As can be observed in Fig.\ 1 of Ref.\
\cite{nielsen04} the strengths of hyperinflation given by the price
index and the exchange rate are quite similar (as in the case of
Weimar Germany treated above).

The Yugoslav hyperinflation has been studied in a number of papers
by using methods developed in the framework of economics. For
instance, we can mention the papers of Petrovi\'c and Mladenovi\'c
\cite{petrovic00} and Nielsen \cite{nielsen04}. These authors consider
the price indexes for 1993:12 and 1994:01 to be unreliable and choose
end their analyses end at the latest 1993:11. This is in line with
standard studies of hyperinflation that mostly ignore the last few
observations. Such a procedure is mainly due to the fact that these
models do not contain structural information over the divergence at
the end of a hyperinflation and big values of price index do not match
into the systematics. However, in that studies the trend of the first
logarithmic differences is analyzed. For instance, Nielsen
\cite{nielsen04} on the basis of his Fig.\ 1 states that such
differences exhibit an exponential growth indicating an accelerating
inflation, but the last three differences are not included in that
figure either.

In standard economic theory, at equilibrium, money determines price
level and implies equilibrium in markets for other variables of the
system. Usually inflation is associated with money supply growth.
However, for instance, there are studies suggesting that the
Yugoslavian hyperinflation (1991$-$4) was not generated by excess
expansion of money stock but, instead, money was accommodating in
this period \cite{joselius02}.

\begin{figure}
\centering\includegraphics[width=8cm, height=7cm]{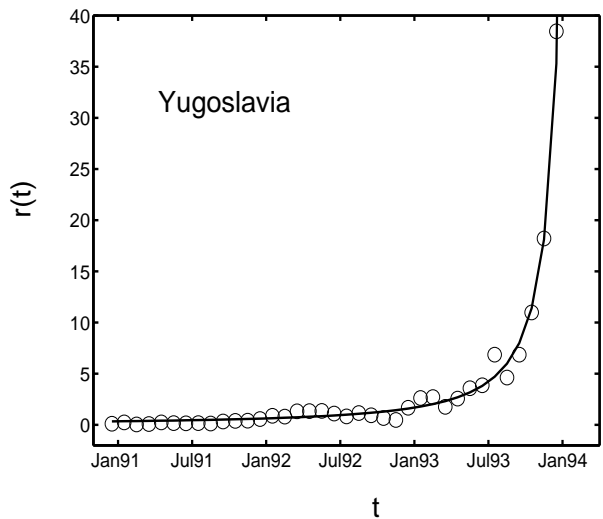}
\caption{\label{fig:Yug_growth} Growth rate of the monthly price index
during the hyperinflation of Yugoslavia marked with open circles.
The solid curve was evaluated by using Eq. (\ref{rate46}) with the
parameters listed in Table \ref{tab:table1}. The vertical line
indicates the critical time $t_c$.}
\end{figure}

In our study we included all the values of price index displayed in
Fig.\ \ref{fig:Yugoslavia}. A fit of these data to Eq.\
(\ref{price11}) yielded the free parameters quoted in Table
\ref{tab:table1}. The value of $\chi$ indicates that the quality of
the fit is good. The uncertainties in the parameters determined for
$\Delta i(t)=25\%$ are smaller than that obtained previously for
Germany and Greece. This is due to the fact that the measured price
index reached very large values determining the position of the
singularity much better than in the former cases. 

In Fig.\ \ref{fig:Yug_growth} we plotted the growth rate $r(t)$
evaluated according to Eq.\ (\ref{rate1}), i.e., by calculating the
first differences of the natural logarithm of the price index. These
data are reproduced very well by Eq.\ (\ref{rate45}) cast into the
form
\begin{equation}
r(t)=C_0\,\Delta t\,\left(\frac{t_c-t_0}{t_c-t}\right)^{1 + \alpha}
\;, \label{rate46}
\end{equation}
and computed with the parameters listed in Table \ref{tab:table1}. A
similar plot was also given by Nielsen \cite{nielsen04}, however, he
did not include the last three data.

We must emphasize that the present model is able to reproduce the
whole series of measured data. The last observations match very well
into the general trend towards the ``explosion''. The obtained $t_c$
predicts a crash at the beginning of 1994:03. The hyperinflation was
stopped with a successful complete currency reform at the end of
1994:01.

\begin{figure}
\centering\includegraphics[width=8cm, height=7cm]{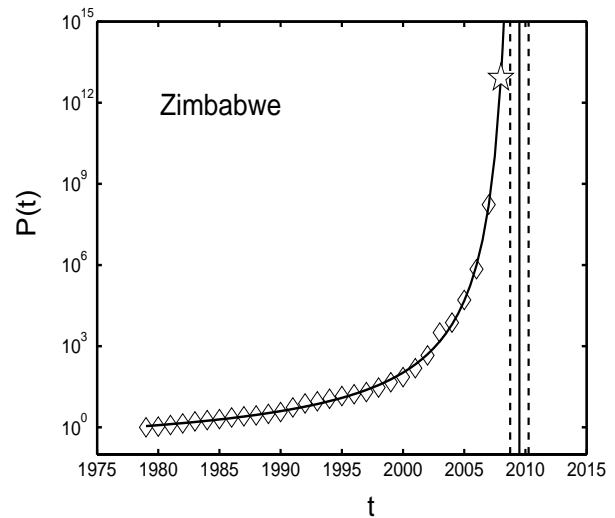}
\caption{\label{fig:Zimbabwe} Semi-logarithmic plot of the yearly
price index of Zimbabwe from 1979 to 2007 marked with open diamonds.
Data are normalized to $P(t_0=1979)=1$. The solid curve is the fit to
Eq.\ (\ref{price11}). The star is the predicted value for 2008. The
vertical lines indicate the obtained critical time $t_c$ and its
error quotes.}
\end{figure}

\subsection{The current tragic case of Zimbabwe}
\label{sec:Zimbabwe}

In 1963 Southern Rhodesia (also known as Rhodesia) chose to remain
a colony when its two partners (Zambia and Malawi) voted for
independence. The country achieved independence on 17 April 1980,
under the name Zimbabwe. At that time the Zimbabwe dollar (ZW-Dollar,
the official symbol is ZWD) was worth about $1.50$ US dollar. Since
the beginning there was a persistent but moderate structural
inflation. Figure \ref{fig:Zimbabwe} shows the yearly price index
taken from files of the Central Statistical Office (CSO) and the
Reserve Bank of Zimbabwe (RBZ).

By looking at Fig.\ \ref{fig:Zimbabwe} one may realize that an
important acceleration of the price index started with the beginning
of the new century. This behavior appeared after the Zimbabwean
government proceeded to finance: i) the expenditure to pay the war
veterans gratitudes in 1997; ii) the intervention in the Democratic
Republic of Congo's war in 1998; and iii) the expenses of a program
of land reforms based on the redistribution of properties in 2000.
The latter undertaking, in practice, led to a weakling of the
agricultural industry and this in turn produced a fall of export
revenues. The resulting deficit was covered by seigniorage. In this
way Zimbabwe started to experience hyperinflation
\cite{makochekanwa07}.

Annual inflation reached about $600\%$ in 2003:12, then fell back to
low triple digits in 2004 before rising again to $600\%$ at end of
2005. In 2006:02, the RBZ announced that the government had printed
$2.1 \times 10^{13}$ ZW-Dollar in order to buy foreign currency to
pay off IMF arrears. In 2006:04 the year-to-year inflation reached
above 1,000{\%}. In 2006:08, the Zimbabwean government issued new
currency slashing three zeros, 1 new Dollar was exchanged for 1,000
old Dollars. The highest denomination was then 1000 new Dollars. The
new money did not provide relief from record inflation. Surging to a
new high above 2,000{\%} in 2007:03 and in 2007:06 rose 7,251{\%}.
Price rises have sharply accelerated in recent months. The CSO stopped
providing data on inflation in October saying that key goods are not
available in stores. In other words, this is a recognition that the
black market already dominates the domestic economy. The denomination
of the largest currency notes is increasing steadily, in 2007:12 the
RBZ unveiled new currency notes with denomination as high as 750,000
ZW-Dollars. In a memo sent at the end of December to financial
institutions to help them close their 2007 books, the RBZ communicated
that the estimated inflation over the past 12 months has totaled
24,059$\%$.

The Consumer Council of Zimbabwe and other observers questioned
whether the figures provided officially reflected the true cost of
living. They stated that real figure is almost certainly much larger.
Recent estimates of Zimbabwean inflation by independent economists
have tended to put it substantially higher ranging from 50,000$\%$
to 100,000$\%$. In any case, the Zimbabwe's inflation is already the
highest in the world and has reached that of Latin America's in the
1980's. For instance, compare with the data of Peru plotted in Fig.\
\ref{fig:Peru}.

The high inflation makes transacting in ZW-Dollars pointless. Indeed
the RBZ has already confirmed that certain farmers will receive
US-Dollar prices for their crops. The severely devalued currency is
also causing many organizations to favor using the US-Dollar instead
of ZW-Dollar. In actual fact, Zimbabwe is closely tracking Germany's
Weimar Republic in the early 1920's. Indeed, there is a close parallel
between the evolution of exchange rate of the ZW-Dollar in the period
2005-2007 and that of German Marks in 1921-1923. Nowadays, the
official exchange rate is 30,739 new ZW-Dollars to a single US-Dollar.
A total devaluation since 1980 amounts about $5 \times 10^7$ a value
similar to the cumulated price index plotted in Fig.\
\ref{fig:Zimbabwe}. Therefore, it is expected that some form of
``US-dollarization'' will establish itself in Zimbabwe in the near
future. This behavior would be in line with previous experiences. It
should be remained that during the hyperinflation many monetary
transactions in Yugoslavia were actually taking place in German Marks
rather than local currency.

A fit of the yearly price index to Eq.\ (\ref{price11}) yielded the
values of the free parameters and $\chi$ quoted in Table
\ref{tab:table1}. The good quality of the adjustment is depicted in
Fig.\ \ref{fig:Zimbabwe}. The predicted critical time $t_c=2009.5$
indicates that the finite-time singularity, in other words the
economic explosion, will occur at mid-2009. By using the obtained
parameters we calculated the price index for the end of 2008, the
result $8.26 \times 10^{12}$ is marked by a star in Fig.\
\ref{fig:Zimbabwe}. This value yields $5 \times 10^6\,${\%} for the
year-to-year inflation. The uncertainties were estimated by assuming
a quote of $25\%$ for the error of measured inflation. The obtained
error bars for $t_c$ are also displayed in Fig.\ \ref{fig:Zimbabwe}.

Zimbabwe's hyperinflation is spiraling to unknown places and could
cause the country's economy to completely collapse within two years.
Due to continued runaway inflation, the RBZ released into circulation
in 2008:01 three very high denomination bearer checks: 1 million ZWD,
5 million ZWD and 10 million ZWD (look at Ref.\ \cite{chao}), these
banknotes will hold tender until July 2008. The latter is worth about
$330$ USD at official exchange rate, but only $3$ USD in the black
market. The Olivera-Tanzi effect is already present. Economic
prospects are bleak, reports of extreme shortages of basic foodstuffs,
fuel, and medical supplies abound. Moreover, such prospects also
indicate that gross domestic product will continue to contract in
2008. Unemployment is around 80 percent and political unrest is
growing (the next presidential elections are scheduled for 2008). In
summary, this is the worst economic crisis since independence from
Britain in 1980.

Therefore, in order to avoid a crash in the near future the government
of Zimbabwe should introduce as soon as possible fundamental reforms
in the economy.

\begin{figure}
\centering\includegraphics[width=8cm, height=7cm]{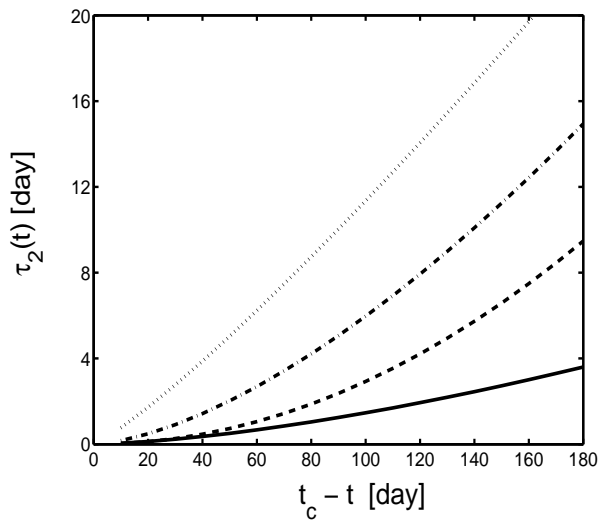}
\caption{\label{fig:tau2} Time interval needed for doubling the price
index in the most severe episodes of hyperinflation as a function of
the time before the crash. The solid curve stands for Yugoslavia, the
dashed curve for Hungary, the dot-dashed curve for Germany, and the
dotted curve for Greece.}
\end{figure}

\subsection{Comparison of severe episodes of hyperinflation}
\label{sec:Comparison}

In order to compare the most severe cases we evaluated the time
$\tau_2(t)$ required to double the price index by using Eq.\
(\ref{tau2}). In this category we included the episodes of Germany,
Greece, Yugoslavia, and Hungary. In the latter case the values of
$\alpha$ and $B$ determined by Sornette {\it et al.} \cite{sornette03}
were used, these data are also quoted in Table \ref{tab:table2}. The
results for the last 180 days previous to the singularity are plotted
in Fig.\ \ref{fig:tau2}. This comparison indicates that at the end of
the cycles Yugoslavia and Hungary suffered the worst effects.

\section{Summary}
\label{sec:summary}

A study of a few important episodes of hyperinflation is reported.
The applied formulation is basically that of Sornette {\it et al.}
\cite{sornette03}. One difference is that in the present work the
expression for cumulated price index is written in terms of free
parameters of the model instead of introducing combinations of them
like quantities $A$ and $B$ defined in Eqs.\ (\ref{Asor}) and
(\ref{Bsor}). In this way the parameters preserve their own physical
meaning. Moreover, the initial time $t_0$ is explicitly retained in
the formulas to avoid any misunderstanding. Another difference is
the implementation of a procedure for estimating uncertainties of the
utilized free parameters. Since the coefficients $A$ and $B$ of Ref.\
\cite{sornette03} are combinations of basic parameters their
uncertainties are correlated, the error analysis in such a case would
be much more complicated.

In the present model the growth rate is accelerating such that the
market price is growing as a power law towards a spontaneous
singularity. This behavior is determined by ``adaptive expectations''
of the people. The effect of other quantities of standard economic
theories does not contribute explicitly. It is worthwhile to notice
that whole cycles are described by fixed values of the exponent
$\gamma$, which preserve the model from possible underestimation of
expectations.

On the other hand, it is well-known that nature does not have pure
singularities in the mathematical sense of the term. Such critical
points are always rounded off or smoothed out by the existence of
friction and dissipation and by the finiteness of the system. This
is a well-known feature of critical points \cite{cardy}. Finite-time
singularities are similarly rounded-off by frictional effects given
in the case of a hyperinflation by currency and economic reforms.

The data of Peru were analyzed to check our procedures. Therefore,
the error analysis is reported in detail.
The study of the hyperinflation of Germany indicates that when there
are reliable data then the analyses of series of price index and
exchange rate lead to very similar results. This behavior confirms
that foreign exchange rate of hyperinflation is positively correlated
with the country's hyperinflation trend. However, once the episode is
finished it is important to reestablish exchange flexibility to allow
reactions to local conditions.
 
We showed that the very extreme cases of Greece and Yugoslavia can be
well described by the present formalism. It should be emphasized that 
both these examples belong to the worst episodes in the hyperinflation
category, where the people expectations were directly influenced by
catastrophic situations described above. For completeness, we also
report a comparison of the most severe cases by determining the time
required to double the price index. 

The last example is, hitherto, rather similar to the Latin America's
cases of the 80's. However, according to our quantitative results the
government of Zimbabwe should perform very deep changes in the
economic system within one year in order to prevent very dreadful
consequences for the society.

Let us finish emphasizing that these lessons should not be lost,
but instead should be kept in mind to avoid the repetition of that
unpleasing experiences. Moreover, one should always remain the
statement of Keynes \cite{keynes30}, namely that: ``even the weakest
government can enforce inflation when it can enforce nothing else''.

\begin{acknowledgments}
This work was supported in part by the Ministry of Culture and
Education of Argentina through Grants CONICET PIP No. 5138/05, 
ANPCyT BID 1728/OC - PICT No. 31980, and UBACYT No. X298.
\end{acknowledgments}

\widetext

\end{document}